\pdfoutput=1

\documentclass[12pt,a4paper]{article}

\usepackage{ifthen} 
\newboolean{pdflatex}
\setboolean{pdflatex}{true} 

\newboolean{articletitles}
\setboolean{articletitles}{true} 

\newboolean{uprightparticles}
\setboolean{uprightparticles}{false} 

\newboolean{inbibliography}
\setboolean{inbibliography}{false} 

\usepackage{xspace}

\usepackage{microtype}
\usepackage{lineno}  
\usepackage{xspace} 
\usepackage{caption} 

\usepackage{graphicx}  
\usepackage{color}
\usepackage{colortbl}
\graphicspath{{./figs/}} 

\usepackage{amsmath} 
\usepackage{amssymb}
\usepackage{amsfonts}
\usepackage{upgreek} 

\newcommand*\patchAmsMathEnvironmentForLineno[1]{%
\expandafter\let\csname old#1\expandafter\endcsname\csname #1\endcsname
\expandafter\let\csname oldend#1\expandafter\endcsname\csname
end#1\endcsname
 \renewenvironment{#1}%
   {\linenomath\csname old#1\endcsname}%
   {\csname oldend#1\endcsname\endlinenomath}%
}
\newcommand*\patchBothAmsMathEnvironmentsForLineno[1]{%
  \patchAmsMathEnvironmentForLineno{#1}%
  \patchAmsMathEnvironmentForLineno{#1*}%
}
\AtBeginDocument{%
\patchBothAmsMathEnvironmentsForLineno{equation}%
\patchBothAmsMathEnvironmentsForLineno{align}%
\patchBothAmsMathEnvironmentsForLineno{flalign}%
\patchBothAmsMathEnvironmentsForLineno{alignat}%
\patchBothAmsMathEnvironmentsForLineno{gather}%
\patchBothAmsMathEnvironmentsForLineno{multline}%
\patchBothAmsMathEnvironmentsForLineno{eqnarray}%
}

\usepackage{hyperref}    
\usepackage[all]{hypcap} 

\def\lhcb {\mbox{LHCb}\xspace}

\def\MagUp {\mbox{\em Mag\kern -0.05em Up}\xspace}

\ifthenelse{\boolean{uprightparticles}}%
{

 \def\Ppi         {\ensuremath{\uppi}\xspace}

 \def\PDelta      {\ensuremath{\Delta}\xspace}                 
 \def\PXi      {\ensuremath{\Xi}\xspace}                 
 \def\PLambda      {\ensuremath{\Lambda}\xspace}                 
 \def\PSigma      {\ensuremath{\Sigma}\xspace}                 
 \def\POmega      {\ensuremath{\Omega}\xspace}                 
 \def\PUpsilon      {\ensuremath{\Upsilon}\xspace}                 
 
 \def\PB      {\ensuremath{\mathrm{B}}\xspace}                 
                  
 \def\PD      {\ensuremath{\mathrm{D}}\xspace}

 \def\PK      {\ensuremath{\mathrm{K}}\xspace}

 \def\Pb      {\ensuremath{\mathrm{b}}\xspace}                 
 \def\Pc      {\ensuremath{\mathrm{c}}\xspace}

 \def\Pi      {\ensuremath{\mathrm{i}}\xspace}

}
{

 \def\Ppi         {\ensuremath{\pi}\xspace}

 \mathchardef\PDelta="7101
 \mathchardef\PXi="7104
 \mathchardef\PLambda="7103
 \mathchardef\PSigma="7106
 \mathchardef\POmega="710A
 \mathchardef\PUpsilon="7107
                  
 \def\PB      {\ensuremath{B}\xspace}                 
                  
 \def\PD      {\ensuremath{D}\xspace}

 \def\PK      {\ensuremath{K}\xspace}

 \def\Pb      {\ensuremath{b}\xspace}                 
 \def\Pc      {\ensuremath{c}\xspace}

 \def\Pi      {\ensuremath{i}\xspace}

}

\makeatletter
\ifcase \@ptsize \relax
  \newcommand{\miniscule}{\@setfontsize\miniscule{4}{5}}
\or
  \newcommand{\miniscule}{\@setfontsize\miniscule{5}{6}}
\or
  \newcommand{\miniscule}{\@setfontsize\miniscule{5}{6}}
\fi
\makeatother

\DeclareRobustCommand{\optbar}[1]{\shortstack{{\miniscule (\rule[.5ex]{1.25em}{.18mm})}
  \\ [-.7ex] $#1$}}

\def\cquark    {{\ensuremath{\Pc}}\xspace}

\def\bquark    {{\ensuremath{\Pb}}\xspace}

\def\pion   {{\ensuremath{\Ppi}}\xspace}

\def\pip    {{\ensuremath{\pion^+}}\xspace}
\def\pim    {{\ensuremath{\pion^-}}\xspace}

\def\kaon    {{\ensuremath{\PK}}\xspace}
  \def\Kbar    {{\kern 0.2em\overline{\kern -0.2em \PK}{}}\xspace}

\def\KorKbar    {\kern 0.18em\optbar{\kern -0.18em K}{}\xspace}

\def\Kp      {{\ensuremath{\kaon^+}}\xspace}
\def\Km      {{\ensuremath{\kaon^-}}\xspace}

\def\KS      {{\ensuremath{\kaon^0_{\rm\scriptscriptstyle S}}}\xspace}

  \def\Dbar    {{\kern 0.2em\overline{\kern -0.2em \PD}{}}\xspace}
\def\D       {{\ensuremath{\PD}}\xspace}

\def\DorDbar    {\kern 0.18em\optbar{\kern -0.18em D}{}\xspace}
\def\Dz      {{\ensuremath{\D^0}}\xspace}
\def\Dzb     {{\ensuremath{\Dbar{}^0}}\xspace}

\def\Dstar   {{\ensuremath{\D^*}}\xspace}

\def\Dstarp  {{\ensuremath{\D^{*+}}}\xspace}
\def\Dstarm  {{\ensuremath{\D^{*-}}}\xspace}

\def\Bbar    {{\ensuremath{\kern 0.18em\overline{\kern -0.18em \PB}{}}}\xspace}

\def\BorBbar    {\kern 0.18em\optbar{\kern -0.18em B}{}\xspace}

  \def\Y#1S{\ensuremath{\PUpsilon{(#1S)}}\xspace}

\def\Lbar        {{\ensuremath{\kern 0.1em\overline{\kern -0.1em\PLambda}}}\xspace}
\def\LorLbar    {\kern 0.18em\optbar{\kern -0.18em \PLambda}{}\xspace}

\newcommand{\decay}[2]{\ensuremath{#1\!\to #2}\xspace}         

\def\to                 {\ensuremath{\rightarrow}\xspace}

\def\order   {{\ensuremath{\mathcal{O}}}\xspace}

\def\CP                {{\ensuremath{C\!P}}\xspace}

\def\AT#1     {\ensuremath{A_{\mathrm{T}}^{#1}}\xspace}           

\def\C#1      {\ensuremath{\mathcal{C}_{#1}}\xspace}                       
\def\Cp#1     {\ensuremath{\mathcal{C}_{#1}^{'}}\xspace}                    
\def\Ceff#1   {\ensuremath{\mathcal{C}_{#1}^{\mathrm{(eff)}}}\xspace}        
\def\Cpeff#1  {\ensuremath{\mathcal{C}_{#1}^{'\mathrm{(eff)}}}\xspace}       
\def\Ope#1    {\ensuremath{\mathcal{O}_{#1}}\xspace}                       
\def\Opep#1   {\ensuremath{\mathcal{O}_{#1}^{'}}\xspace}                    

\newcommand{\tev}{\ifthenelse{\boolean{inbibliography}}{\ensuremath{~T\kern -0.05em eV}\xspace}{\ensuremath{\mathrm{\,Te\kern -0.1em V}}}\xspace}
\newcommand{\gev}{\ensuremath{\mathrm{\,Ge\kern -0.1em V}}\xspace}
\newcommand{\mev}{\ensuremath{\mathrm{\,Me\kern -0.1em V}}\xspace}
\newcommand{\kev}{\ensuremath{\mathrm{\,ke\kern -0.1em V}}\xspace}
\newcommand{\ev}{\ensuremath{\mathrm{\,e\kern -0.1em V}}\xspace}
\newcommand{\gevc}{\ensuremath{{\mathrm{\,Ge\kern -0.1em V\!/}c}}\xspace}
\newcommand{\mevc}{\ensuremath{{\mathrm{\,Me\kern -0.1em V\!/}c}}\xspace}
\newcommand{\gevcc}{\ensuremath{{\mathrm{\,Ge\kern -0.1em V\!/}c^2}}\xspace}
\newcommand{\gevgevcccc}{\ensuremath{{\mathrm{\,Ge\kern -0.1em V^2\!/}c^4}}\xspace}
\newcommand{\mevcc}{\ensuremath{{\mathrm{\,Me\kern -0.1em V\!/}c^2}}\xspace}

\def\m    {\ensuremath{\rm \,m}\xspace}

\def\mum  {\ensuremath{{\,\upmu\rm m}}\xspace}

\def\invfb   {\ensuremath{\mbox{\,fb}^{-1}}\xspace}

\def\order{{\ensuremath{\cal O}}\xspace}

\def\gsim{{~\raise.15em\hbox{$>$}\kern-.85em
          \lower.35em\hbox{$\sim$}~}\xspace}
\def\lsim{{~\raise.15em\hbox{$<$}\kern-.85em
          \lower.35em\hbox{$\sim$}~}\xspace}

\def\ptot       {\mbox{$p$}\xspace}
\def\pt         {\mbox{$p_{\rm T}$}\xspace}

\newcommand{\lum} {\ensuremath{\mathcal{L}}\xspace}

\def\evtgen     {\mbox{\textsc{EvtGen}}\xspace}

\def\geant      {\mbox{\textsc{Geant4}}\xspace}

\def\photos     {\mbox{\textsc{Photos}}\xspace}

\def\pythia     {\mbox{\textsc{Pythia}}\xspace}

\def\tell1  {TELL1\xspace}
\def\ukl1   {UKL1\xspace}

\newcommand{\eg}{\mbox{\itshape e.g.}\xspace}

\usepackage{cite} 
\usepackage{mciteplus}

\begin{document}

\newcommand{\CPV}{\CP violation\xspace}
\newcommand{\CPA}{\CP asymmetry\xspace}
\newcommand{\CPAs}{\CP asymmetries\xspace}
\newcommand{\CPVh}{\CP-violation\xspace}
\newcommand{\CPAh}{\CP-asymmetry\xspace}
\newcommand{\mydec}{\decay{\Dz}{\KS\KS}}
\newcommand{\mydecBar}{\decay{\Dzb}{\KS\KS}}
\newcommand{\myDstarDec}{\decay{\Dstarp}{\Dz\pip}}
\newcommand{\myControlDec}{\decay{\Dz}{\Km\pip}}
\newcommand{\ts}{technical stop\xspace}
\newcommand{\Ts}{technical stop\xspace}
\newcommand{\nuTwo}{\ensuremath{{\nu_2^\prime}}\xspace}
\newcommand{\nuThree}{\ensuremath{{\nu_3}}\xspace}
\newcommand{\lnpt}{\ensuremath{{C_{\text{lnpt}}}}\xspace}
\newcommand{\mrange}[3]{\ensuremath{{#1 \in #2 \pm #3}}\xspace}
\newcommand{\asym}{\ensuremath{{\cal A}_{\CP}}\xspace}
\newcommand{\myintlum}[1]{\ensuremath{\int\lum\, dt = #1}\xspace}
\newcommand{\aLL}{LL\xspace}
\newcommand{\aLLtrig}{LLtrig\xspace}
\newcommand{\aLD}{LD\xspace}
\newcommand{\aDD}{DD\xspace}

\renewcommand{\thefootnote}{\fnsymbol{footnote}}
\setcounter{footnote}{1}

\begin{titlepage}
\pagenumbering{roman}

\vspace*{-1.5cm}
\centerline{\large EUROPEAN ORGANIZATION FOR NUCLEAR RESEARCH (CERN)}
\vspace*{1.5cm}
\hspace*{-0.5cm}
\begin{tabular*}{\linewidth}{lc@{\extracolsep{\fill}}r}
\ifthenelse{\boolean{pdflatex}}
{\vspace*{-2.7cm}\mbox{\!\!\!\includegraphics[width=.14\textwidth]{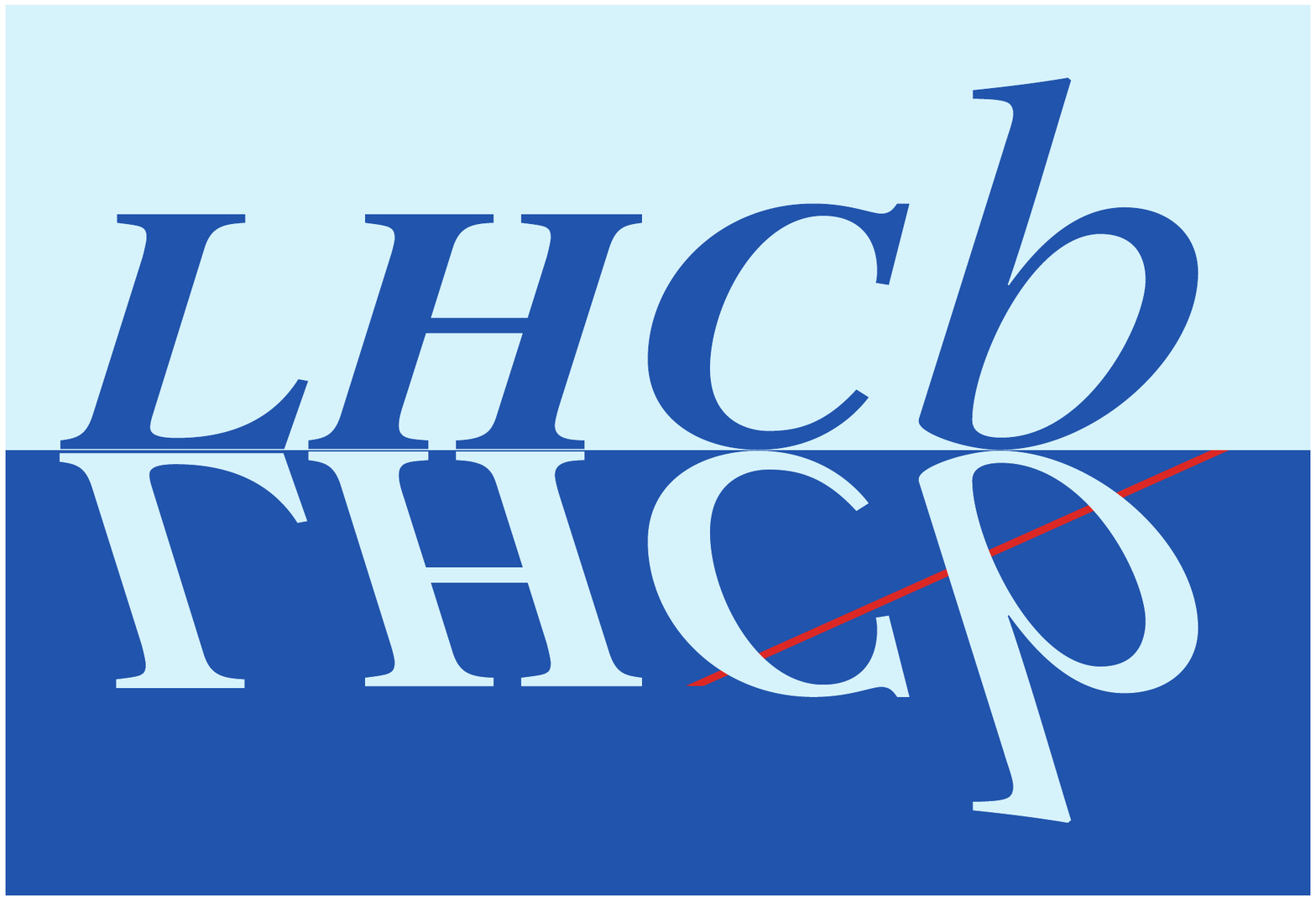}} & &}%
{\vspace*{-1.2cm}\mbox{\!\!\!\includegraphics[width=.12\textwidth]{lhcb-logo.eps}} & &}%
\\
 & & CERN-PH-EP-2015-215 \\  
 & & LHCb-PAPER-2015-030 \\  
 & & Oktober 8, 2015 \\ 
 & & \\
\end{tabular*}

\vspace*{3.0cm}

{\bf\boldmath\huge
\begin{center}
  Measurement of the time-integrated \CPA in \mydec decays
\end{center}
}

\vspace*{2.0cm}

\begin{center}
The LHCb collaboration\footnote{Authors are listed at the end of this paper.}
\end{center}

\vspace{\fill}

\begin{abstract}
  \noindent
The time-integrated \CPA in the decay \mydec is measured 
using $3\invfb$ of proton-proton collision data collected by the 
LHCb experiment at centre-of-mass energies of 7 and 8\tev. The 
flavour of the \Dz meson is determined by use of the decay
\myDstarDec and its charge conjugate mode.
The result is
\[ \asym = -0.029 \: \pm \: 0.052 \: \pm \: 0.022 \ ,\]
where the first uncertainty is statistical and the second systematic. The result is consistent with Standard Model expectations and improves the uncertainty with respect to the only previous measurement of this quantity by more than a factor of three.
\end{abstract}

\vspace*{1.0cm}

\begin{center}
  Accepted by JHEP 
\end{center}

\vspace{\fill}

{\footnotesize 
\centerline{\copyright~CERN on behalf of the \lhcb collaboration, license \href{http://creativecommons.org/licenses/by/4.0/}{CC-BY-4.0}.}}
\vspace*{2mm}

\end{titlepage}


\newpage
\setcounter{page}{2}
\mbox{~}
\newpage

\cleardoublepage

\renewcommand{\thefootnote}{\arabic{footnote}}
\setcounter{footnote}{0}


\pagestyle{plain} 
\setcounter{page}{1}
\pagenumbering{arabic}

\section{Introduction}
\label{sec:Introduction}

In the Standard Model, \CPV in charm decays is expected to be small
and hence potentially sensitive to contributions from New Physics. Although measurements
of the time-integrated \CPA in \Dz decays to pairs of charged mesons 
showed hints of \CPA at the level of 0.7\,\%, the combined results 
are not yet conclusive\cite{Aubert:2007if, Collaboration:2012qw, Ko:2012px, LHCb-CONF-2013-003, LHCb-PAPER-2014-013}. Of particular interest, both for the search of 
New Physics and for the understanding of penguin contributions,
are decays of \Dz mesons into a pair of 
neutral mesons, such as the decay \mydec\cite{hiller, uli}.\footnote{The inclusion of charge-conjugate processes is implied throughout this article.} If the \CPA in \Dz decays to charged mesons is
confirmed and assuming moderate breaking of the SU(3) flavour symmetry, 
the \CPA of this mode could be of \order{(1\,\%)} or even larger\cite{hiller}. From a more
recent Standard Model based analysis of the contributing amplitudes, a 
95\,\% confidence level upper limit of 1.1\,\% for direct \CPV in the decay 
\mydec has been derived\cite{uli}. The single previous measurement gave 
$\asym = (23 \pm 19)$\,\%~\cite{D0KSKSCleo}. 

Here, we present the first result from the LHCb collaboration on \CPV in the decay \mydec. The measurement is based on \Dz mesons
originating from \myDstarDec decays, where the flavour of the \Dz meson
can be inferred from the charge of the ``slow'' pion from the \Dstarp decay. 
Throughout this document, \Dstarp stands for $D^*(2010)^+$. As a 
control channel, the decay \decay{\Dz}{\Km\pip} is used to estimate production 
and detection asymmetries.
The analysis uses $3\invfb$ of proton-proton collision data collected with the \lhcb detector in 2011 and in 2012, at centre-of-mass energies of $7\tev$ and $8\tev$, respectively.

\section{Detector and simulation}
\label{sec:Detector}

The \lhcb detector\cite{Alves:2008zz,LHCb-DP-2014-002} is a single-arm forward
spectrometer covering the \mbox{pseudorapidity} range $2<\eta <5$,
designed for the study of particles containing \bquark or \cquark
quarks. The detector includes a high-precision tracking system
consisting of a silicon-strip vertex detector surrounding the $pp$
interaction region, a large-area silicon-strip detector located
upstream of a dipole magnet with a bending power of about
$4{\rm\,Tm}$, and three stations of silicon-strip detectors and straw
drift tubes placed downstream of the magnet.
The tracking system provides a measurement of momentum, \ptot, of charged particles with
a relative uncertainty that varies from 0.5\% at low momentum to 1.0\% at 200\gev.\footnote{We use natural units where $c = 1$.}
The minimum distance of a track to a primary vertex (PV), the impact parameter, is measured with a resolution of $(15+29/\pt)\mum$,
where \pt is the component of the momentum transverse to the beam, in\,\gev.
The polarity of the dipole magnet is reversed regularly throughout the data-taking period, which allows to determine and correct for
charge asymmetries due to the detector geometry.

Different types of charged hadrons are distinguished using information
from two ring-imaging Cherenkov detectors. 
Photons, electrons and hadrons are identified by a calorimeter system consisting of
scintillating-pad and preshower detectors, an electromagnetic
calorimeter and a hadronic calorimeter. Muons are identified by a
system composed of alternating layers of iron and multiwire
proportional chambers.
The online event selection is performed by a trigger, 
which consists of a hardware stage, based on information from the calorimeter and muon
systems, followed by a software stage, which applies a full event
reconstruction.

In the simulation, $pp$ collisions are generated using
\pythia~\cite{Sjostrand:2006za,*Sjostrand:2007gs} with a specific \lhcb
configuration~\cite{LHCb-PROC-2010-056}.  Decays of hadronic particles
are described by \evtgen~\cite{Lange:2001uf}, in which final state
radiation is generated using \photos~\cite{Golonka:2005pn}. The
interaction of the generated particles with the detector, and its
response, are implemented using the \geant
toolkit~\cite{Allison:2006ve, *Agostinelli:2002hh} as described in
Ref.~\cite{LHCb-PROC-2011-006}.

\section{Selection}
\label{sec:Selection}

Signal decays are reconstructed in the decay mode \myDstarDec with \mydec and \decay{\KS}{\pip\pim}\cite{Hulsbergen:2005pu}. 
To collect as many \mydec decays as possible, events from all available physics triggers are considered. Candidate events are accepted if the four pions assigned to the \KS decays are sufficient to trigger the event or if the rest of the event, without the slow pion from the \Dstar decay, satisfies a trigger condition. Excluding the slow pion from the trigger decision minimises any bias on the \CPA due to the trigger.

The \decay{\KS}{\pip\pim} decays are reconstructed in two different categories:
the first involves \KS mesons that decay early enough for the
daughter pions to be reconstructed in the vertex detector; the
second contains \KS mesons that decay later such that daughter track segments are only reconstructed in the tracking detectors downstream of the vertex detector. These categories are
referred to as \emph{long} (L) and \emph{downstream} (D), respectively. The
less abundant long category has better momentum and vertex resolution than the
downstream category. 
As the final state contains two \KS mesons, there are three possible combinations labeled \aLL, \aLD, and \aDD. 
A dedicated software trigger selection for the \mydec decay was implemented in 2012. 
As this trigger only accepts signal candidates composed of two long \KS, a fourth \Dz category, \aLLtrig, is defined where the dedicated trigger accepted the signal candidate. 
The four categories are listed in Table~\ref{tab:categories}.

\begin{table}[t]
\caption{Definition of candidate categories.}
\begin{center}
\begin{tabular}{ll}
\hline
Categories & Description \\
\hline
\aLL & both \KS are of category long and \emph{not} selected by the dedicated trigger \\
\aLD & one \KS is long, the other one is downstream \\
\aDD & both \KS are downstream\\
\aLLtrig & both \KS are of category long and selected by the dedicated trigger \\
\end{tabular}
\end{center}
\label{tab:categories}
\end{table}%

The decay vertex of the \Dz candidate is reconstructed from the pion trajectories, constraining the \KS mass to its known value\cite{PDG2014} and the \Dz flight direction to point to a PV\cite{Hulsbergen:2005pu}. 
To reduce the contamination from the decay \decay{\Dz}{\KS\pip\pim}, only candidates with significant decay times for both \KS decays are accepted. This requirement has a signal efficiency of over 99\,\%. 
The \Dz candidate is combined with a pion to produce a \Dstar candidate.
Fiducial cuts are applied on the kinematic properties of this slow pion to remove regions where the detection charge asymmetry is large. A cut on the invariant mass of the \KS\KS system of $\pm 20\mev$ around the known value of the \Dz mass is applied. The efficiency of this cut is 98\,\% in all categories, except for \aDD, where it is 94\,\%. Candidates are further selected by requiring the difference between the \Dstarp and \Dz candidate masses, $\Delta m \equiv m_{\Dstar} - m_{\Dz}$, to be less than $155\mev$. 

To further reduce combinatorial background a multivariate analysis (MVA) method\cite{acat08, acat10} is used. It is based on a rule-based learner 
applying the methods of bagging\cite{breimanBagging} and instance weighting (see, \eg, Refs.\cite{ting,zhao}).
Separate samples have been used for training and testing, where simulated events have been used as signal proxy, while the background sample was a mixture of data from \Dz sidebands (the mass ranges $1764.84 - 1844.84\mev$ and $1884.84 - 1964.84\mev$) and simulation.
Kinematic quantities, decay time variables, geometric quantities, and fit quality variables are used as input to the MVA. The selection of variables and the training have been done separately for the different categories.
For the MVA optimisation, the figure of merit $S/\sqrt{S+B}$ is used, where $S$ is the expected number of signal candidates in the signal region defined by a $\pm1.5\mev$ window around the known $\Delta m$ value\cite{PDG2014}, while $B$ is the number of background candidates in the signal region. The optimal points have signal efficiencies of 95\,\%, 51\,\%, 47\,\%, and 37\,\%, and background retentions of 50\,\%, 0.33\,\%, 0.40\,\%, and 0.73\,\% for \aLLtrig, \aLL, \aLD, and \aDD, respectively. The large background retention for the \aLLtrig category is due to the fact that the dedicated trigger intrinsically has a much lower 
background level.
The control channel \myControlDec is selected by cuts on kinematic quantities, vertex quality variables, geometric quantities, and decay time variables. In addition, the same mass requirements and fiducial cuts on the slow pion are applied as in case of the signal channel.
Due to the large number of control channel candidates, which is much larger than needed for this analysis, 1\,\% of candidates are accepted at random.

\section{Asymmetry measurement}
\label{sec:Yields}

The \CPA is obtained as
\begin{equation}
	\asym = \frac{N^+ - N^-}{N^+ + N^-}
\end{equation}
for each category, where $N^+$ ($N^-$) is the yield determined from a fit to the data for a positive (negative) charge of the slow pion. 

The yields are determined from an extended unbinned maximum 
likelihood fit to the $\Delta m$ distribution. In the fit model,
the signal is described by a sum of three Gaussian functions,
where the two narrower ones are required to have the same mean value.
The parameters of the narrowest Gaussian function and the mean value of 
the widest one are allowed to float, while the ratios between widths, as well as those between normalisations,
of all three are constrained to the values 
found in the simulation. The background is parametrised by the 
product of an exponential function and a power law for the phase-space 
threshold at the pion mass
\begin{equation}
  f_{\rm bg} = C_{\rm bg} 
(\Delta m - m_{\pip})^p e^{-(\Delta m - \m_{\pip})\alpha} \;.
\end{equation}
Here $C_{\rm bg}$, $p$ and $\alpha$ are determined by the fit. 
Independent fits are performed for the four categories. In
each category, the background parameters and the shape parameters
of the signal component are shared between the two charges of the
slow pion. 

\begin{figure}
\centering
\includegraphics[width = 0.45\textwidth]{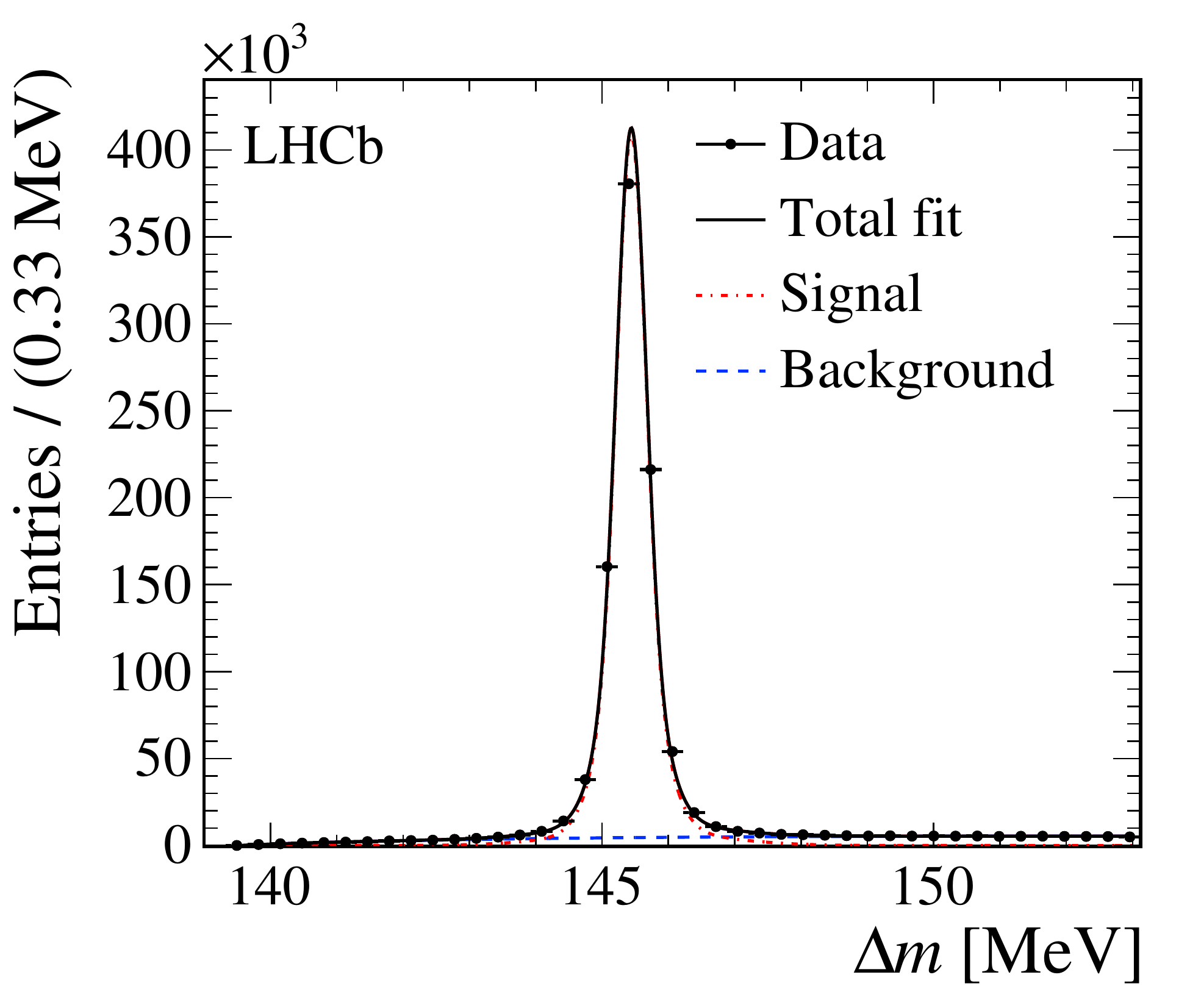}
\includegraphics[width = 0.45\textwidth]{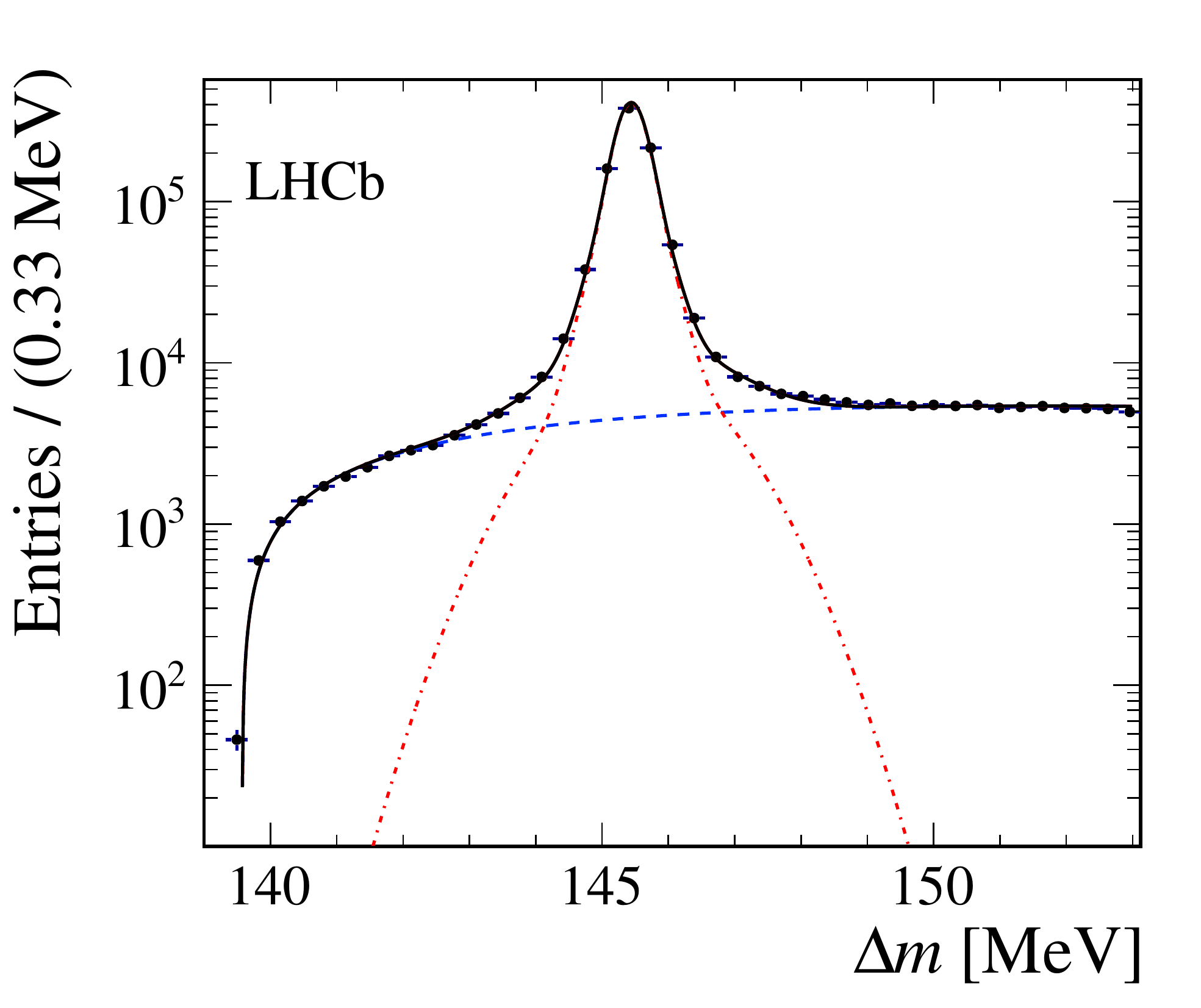}\\
\caption{Distribution of $\Delta m$ for the control channel \myControlDec in (left) linear and in (right) logarithmic scale. The solid (black) line corresponds to the total fit, the dashed (blue) line corresponds to the background, and the dash-dotted (red) line represents the signal contribution.}
\label{fig:massMagAnyCLAllDataCL}
\end{figure}

\begin{figure}
\centering
\includegraphics[width = 0.49\textwidth]{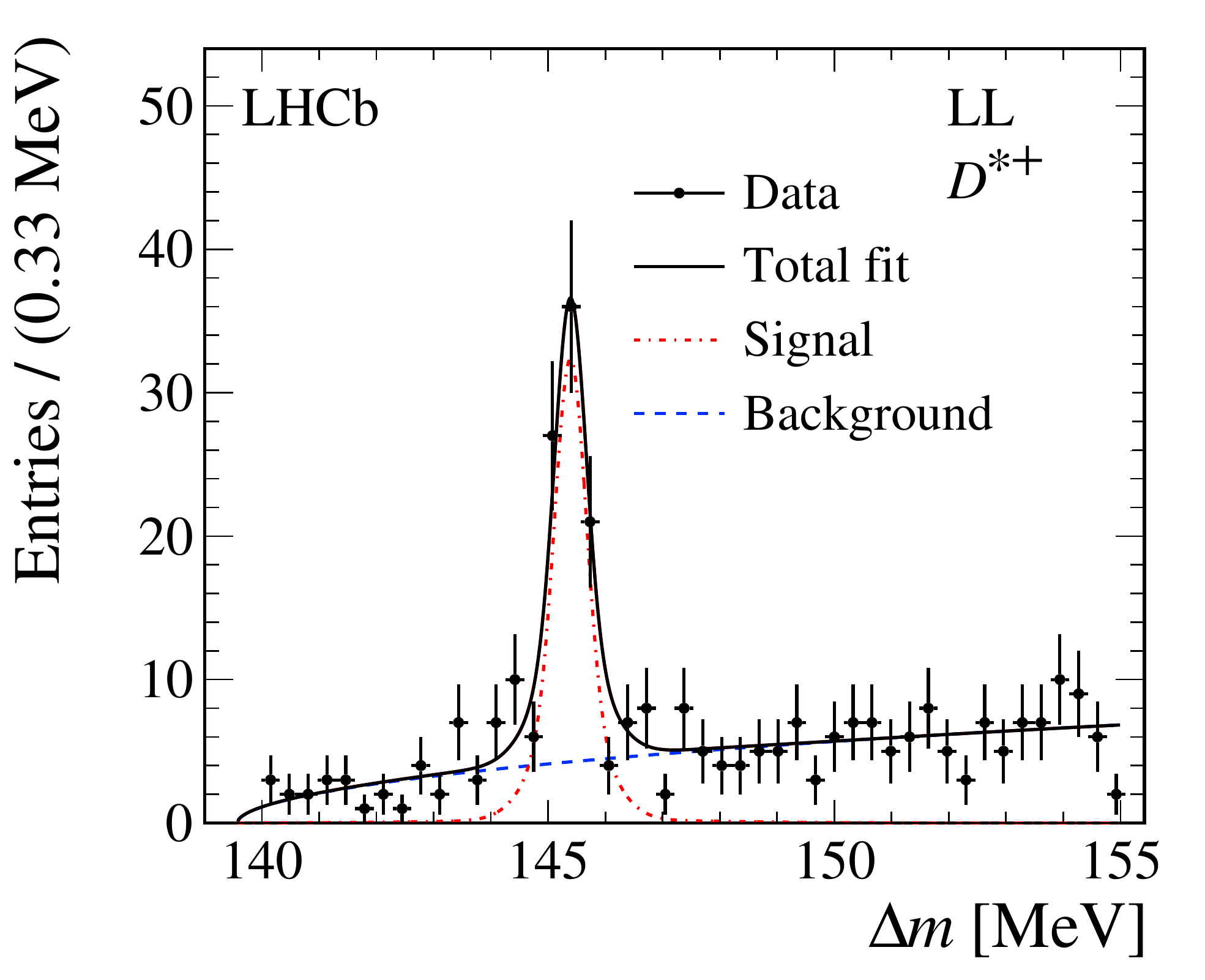}
\includegraphics[width = 0.49\textwidth]{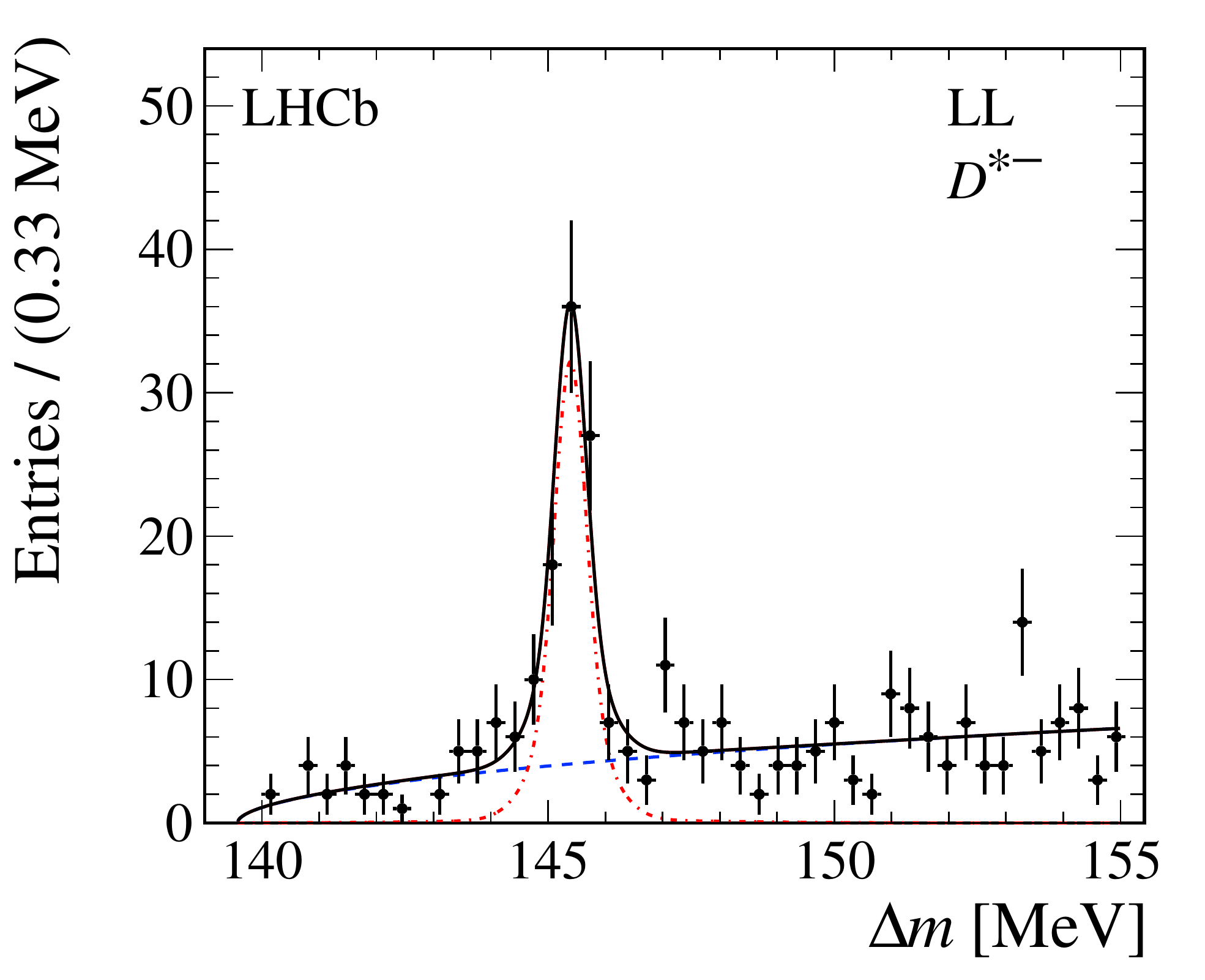}\\
\includegraphics[width = 0.49\textwidth]{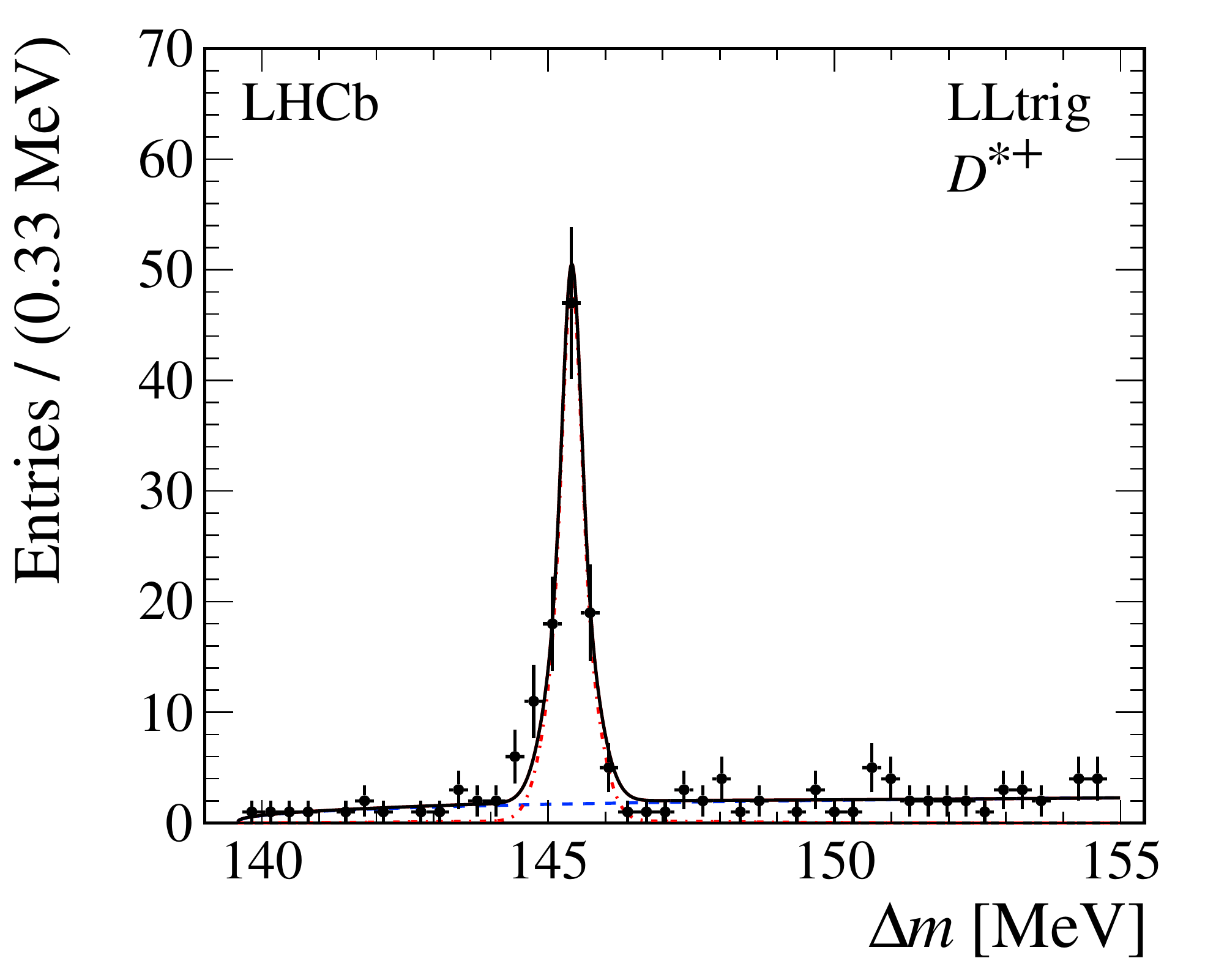}
\includegraphics[width = 0.49\textwidth]{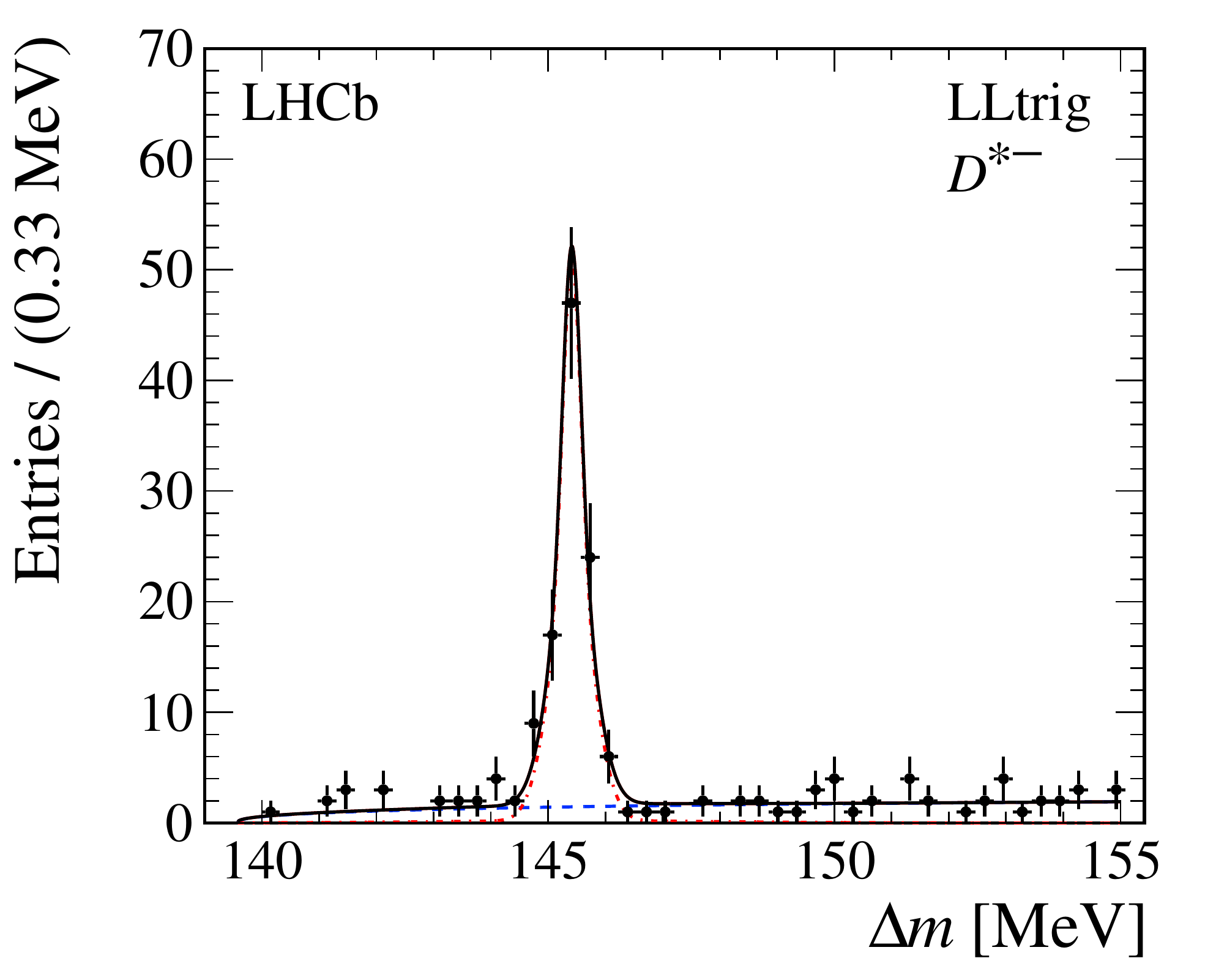}\\
\caption{Distributions of $\Delta m$ split into (left) \Dstarp, (right) \Dstarm and (top) \aLL, (bottom) \aLLtrig, including the fit function. The solid (black) line corresponds to the total fit, the dashed (blue) line corresponds to the background, and the dash-dotted (red) line represents the signal contribution.}
\label{fig:massMagAnyAllDataANpm1}
\end{figure}

\begin{figure}
\centering
\includegraphics[width = 0.49\textwidth]{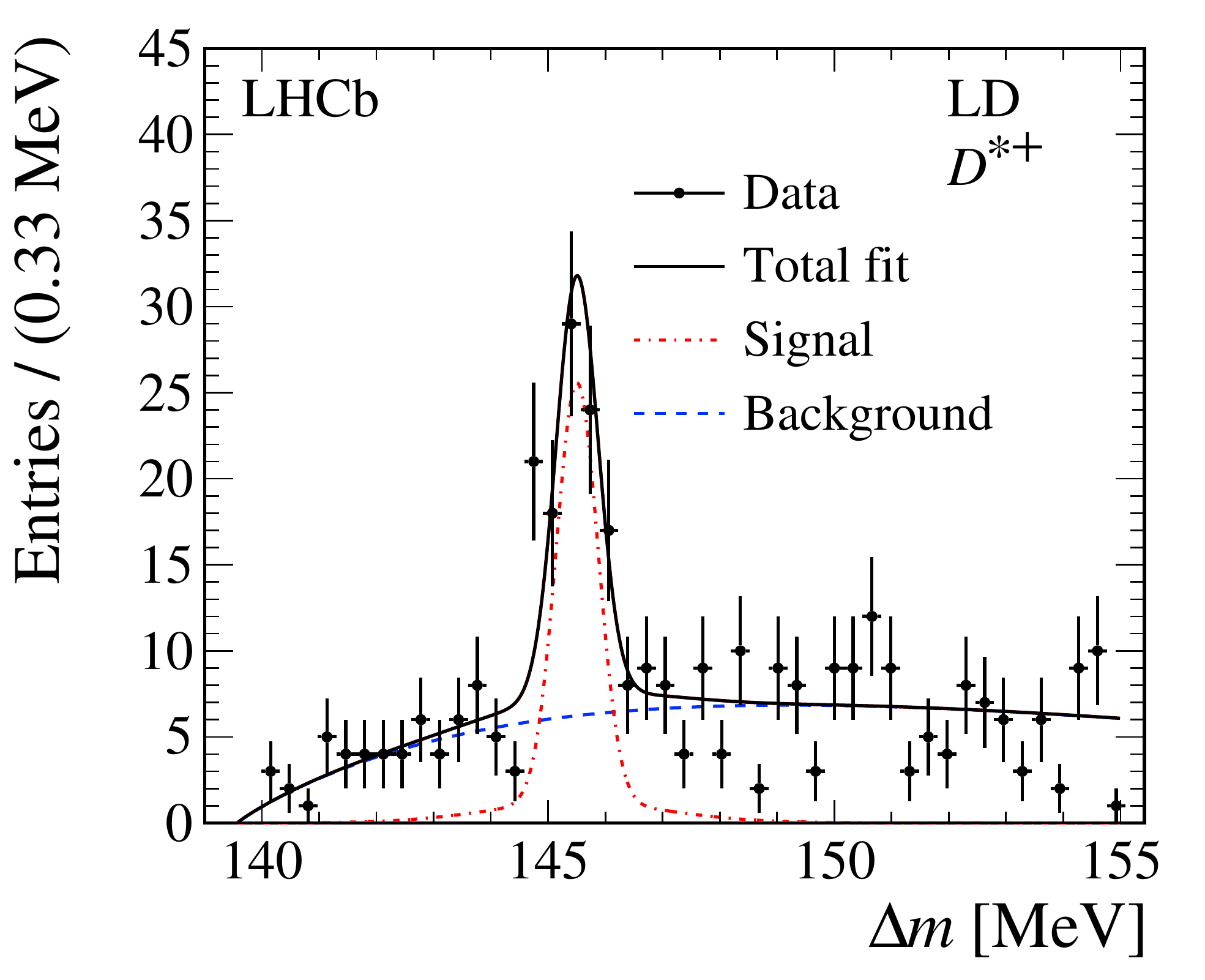}
\includegraphics[width = 0.49\textwidth]{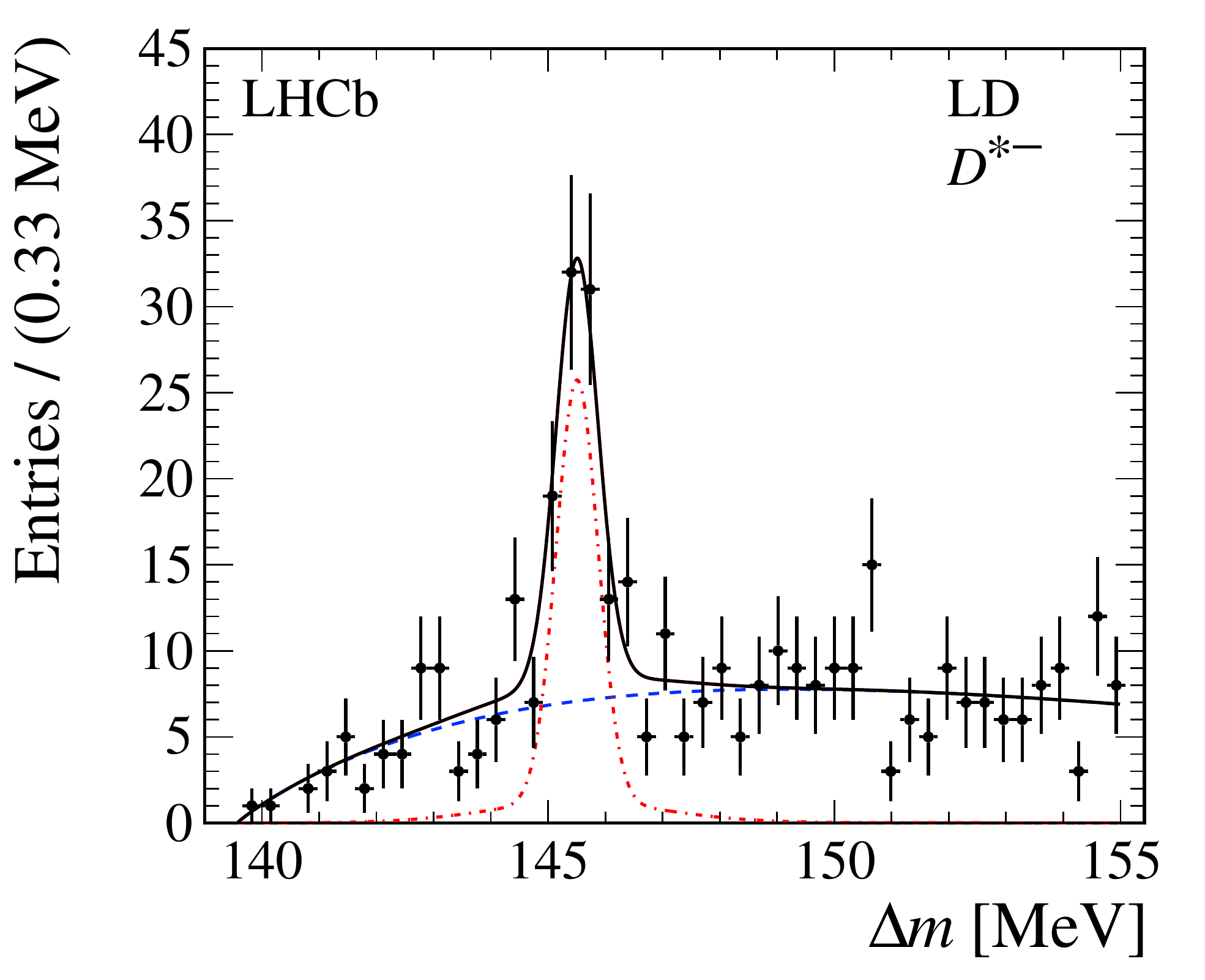}\\
\includegraphics[width = 0.49\textwidth]{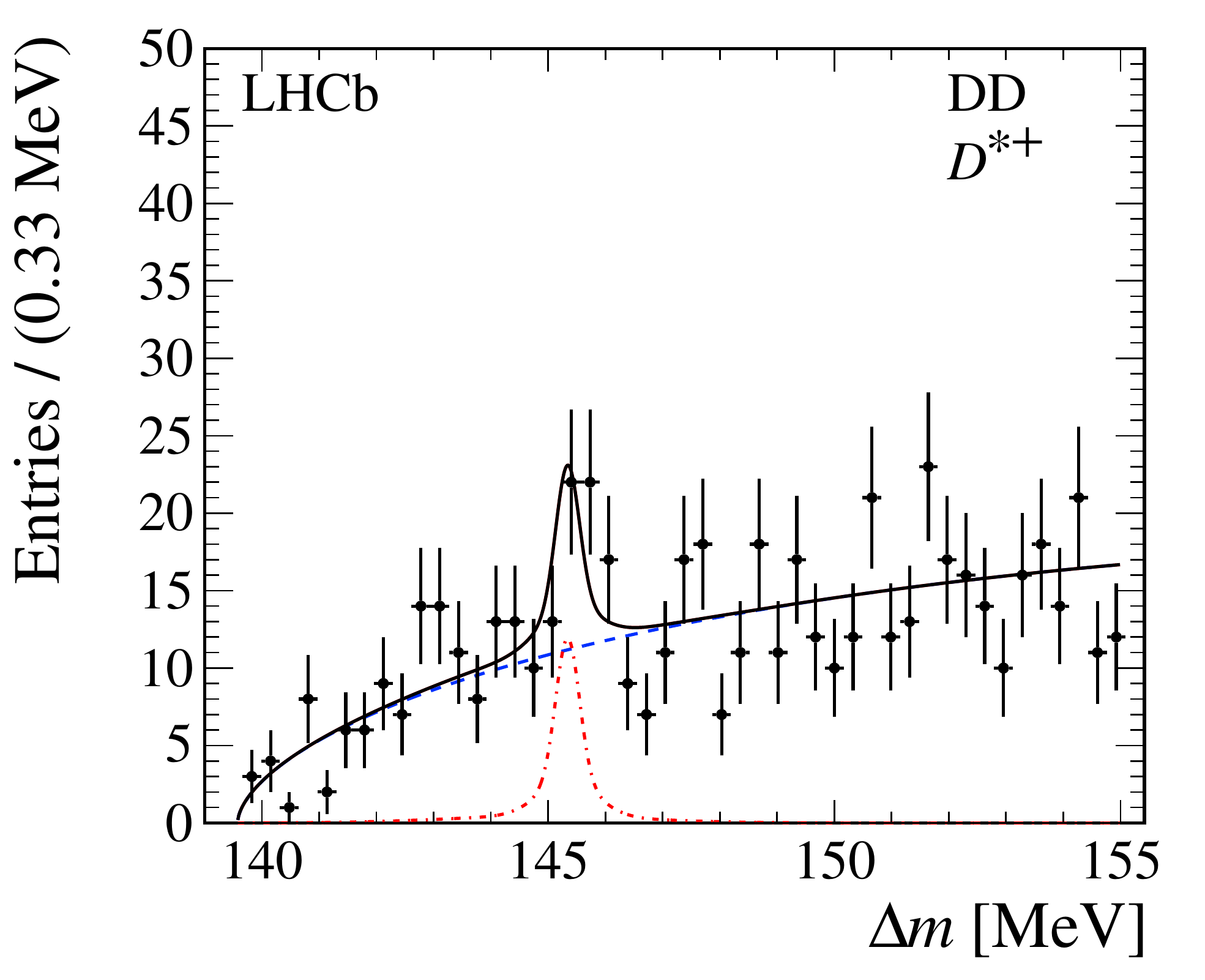}
\includegraphics[width = 0.49\textwidth]{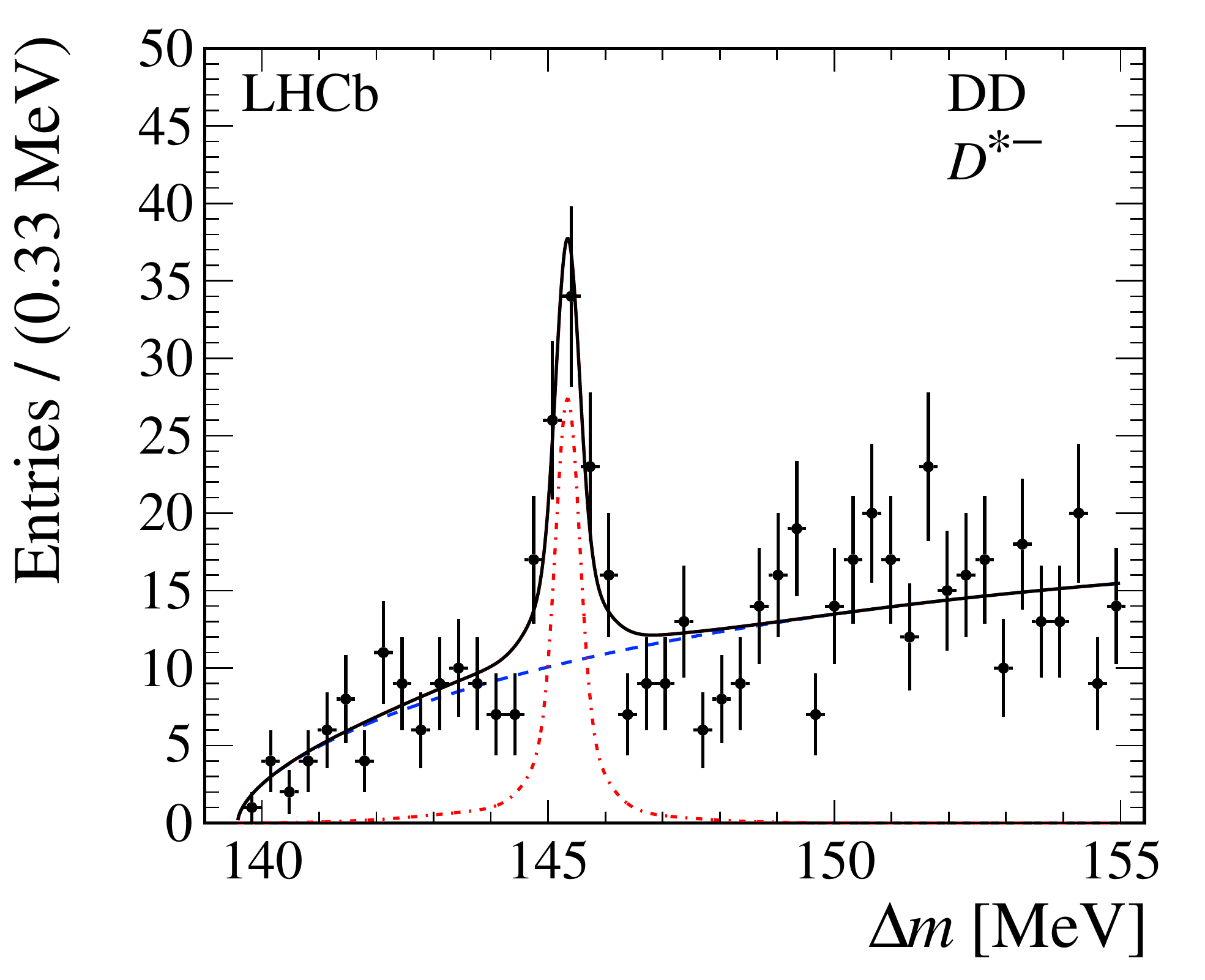}\\
\caption{Distributions of $\Delta m$ split into (left) \Dstarp, (right) \Dstarm and (top) \aLD, (bottom) \aDD, including the fit function. The solid (black) line corresponds to the total fit, the dashed (blue) line corresponds to the background, while the dash-dotted (red) line represents the signal contribution.}
\label{fig:massMagAnyAllDataANpm2}
\end{figure}

The $\Delta m$ distribution for the control channel,
summed over both charges of the slow pion with the fit function overlaid is shown in Fig.\,\ref{fig:massMagAnyCLAllDataCL}. Figures~\ref{fig:massMagAnyAllDataANpm1} and~\ref{fig:massMagAnyAllDataANpm2} show
the $\Delta m$ distributions and the fit for each of the four 
categories and the two slow-pion charges. Table~\ref{tab:yieldResult} lists the 
yields from the nominal fits and the resulting asymmetries.
To obtain the final result, the asymmetries of the four signal categories are combined by taking the weighted mean.

\begin{table}[htp]
\caption{Number of signal candidates and \CP asymmetry obtained from the fits in the four categories.}
\begin{center}
\begin{tabular}{lcccc}
\hline
Category & $N^+$ & $N^-$ & \asym \\
\hline
\aLL & $86 \pm 11$ & $86 \pm 12$ & $\phantom{+}0.00 \pm 0.09$  \\
\aLD & $82 \pm 14$ & $83 \pm 13$ & $-0.00 \pm 0.11$  \\
\aDD & $29 \pm 14$ & $66 \pm 14$ & $-0.39 \pm 0.23$ \\
\aLLtrig & $96 \pm 11$ & $99 \pm 11$ & $-0.02 \pm 0.08$  \\
\hline
combined & & & $-0.029 \pm 0.052$ \\
 \end{tabular}
\end{center}
\label{tab:yieldResult}
\end{table}%

\section{Systematic uncertainties}
\label{sec:syst}

The main sources of systematic effects are due to production and detection
asymmetries and possible biases in the signal extraction method.
The systematic uncertainty related to the signal extraction is
estimated by comparing the nominal fit with an alternative one, 
where outside of the signal region of $\pm 1.5$\,MeV around the 
known $\Delta m$-value, only the background component is 
fitted. The signal yield is obtained by subtracting the
background extrapolated into the signal region from the total observed 
yield. For the combined \CPA a difference of 0.019 is 
found, which is assigned as a systematic uncertainty.

The systematic effects that arise due to the slow pion charge asymmetry in the detector and a possible charge asymmetry of \Dstar production in $pp$ collisions in the \lhcb acceptance are determined using the control channel. 
However, the control channel contains charged kaons which introduce an additional detection asymmetry, as the interaction cross-sections of \Kp and \Km with the detector material are different. In Ref.\cite{LHCB-PAPER-2013-053}, the charged kaon detection asymmetry has been measured to be in the range $0.008$ to $0.012$. Assuming the pion detection asymmetry to be negligible, and including possible trigger effects, a correction of $-0.010 \pm 0.005$ 
is applied to the observed asymmetry in the control channel,
resulting in a corrected value of $-0.009 \pm 0.005$.
The absolute value of this number and its uncertainty are added in quadrature and assigned as a conservative estimate of the systematic uncertainty due to production and detection asymmetries.

Other checks have been performed but
found to have statistically insignificant effects. These tests include the split into different trigger types, different run periods, and different magnet polarities. Also the effect of a possible difference in contamination by charm from beauty decays between signal and the control channel has been checked and found to be negligible.
The total systematic uncertainty is calculated from the quadratic sum of the two dominant contributions, which come from the signal extraction (0.019) and the detection and production asymmetry (0.011), giving 0.022 for the total.

\section{Result}
\label{sec:results}

The time-integrated \CPA in the decay \mydec is determined to be 
\[ \asym = -0.029 \: \pm \: 0.052 \: \pm \: 0.022 \ ,\]
where the first uncertainty is statistical and the second
systematic.
The result is consistent with no \CPV and with Standard Model expectations\cite{uli}. This is the single best measurement of this quantity to date, 
with an uncertainty more than three times smaller than the previous determination\cite{D0KSKSCleo}.

\section*{Acknowledgements}

\noindent We express our gratitude to our colleagues in the CERN
accelerator departments for the excellent performance of the LHC. We
thank the technical and administrative staff at the LHCb
institutes. We acknowledge support from CERN and from the national
agencies: CAPES, CNPq, FAPERJ and FINEP (Brazil); NSFC (China);
CNRS/IN2P3 (France); BMBF, DFG, HGF and MPG (Germany); INFN (Italy); 
FOM and NWO (The Netherlands); MNiSW and NCN (Poland); MEN/IFA (Romania); 
MinES and FANO (Russia); MinECo (Spain); SNSF and SER (Switzerland); 
NASU (Ukraine); STFC (United Kingdom); NSF (USA).
The Tier1 computing centres are supported by IN2P3 (France), KIT and BMBF 
(Germany), INFN (Italy), NWO and SURF (The Netherlands), PIC (Spain), GridPP 
(United Kingdom).
We are indebted to the communities behind the multiple open 
source software packages on which we depend. We are also thankful for the 
computing resources and the access to software R\&D tools provided by Yandex LLC (Russia).
Individual groups or members have received support from 
EPLANET, Marie Sk\l{}odowska-Curie Actions and ERC (European Union), 
Conseil g\'{e}n\'{e}ral de Haute-Savoie, Labex ENIGMASS and OCEVU, 
R\'{e}gion Auvergne (France), RFBR (Russia), XuntaGal and GENCAT (Spain), Royal Society and Royal
Commission for the Exhibition of 1851 (United Kingdom).


\appendix

\addcontentsline{toc}{section}{References}
\setboolean{inbibliography}{true}
\bibliographystyle{LHCb}
\bibliography{d0ksks,main,LHCb-PAPER,LHCb-CONF,LHCb-DP}
\newpage

\newpage
\centerline{\large\bf LHCb collaboration}
\begin{flushleft}
\small
R.~Aaij$^{38}$, 
B.~Adeva$^{37}$, 
M.~Adinolfi$^{46}$, 
A.~Affolder$^{52}$, 
Z.~Ajaltouni$^{5}$, 
S.~Akar$^{6}$, 
J.~Albrecht$^{9}$, 
F.~Alessio$^{38}$, 
M.~Alexander$^{51}$, 
S.~Ali$^{41}$
% ...

{\footnotesize \it
% ...
$ ^{5}$Clermont Universit\'{e}, Universit\'{e} Blaise Pascal, CNRS/IN2P3, LPC, Clermont-Ferrand, France\\
$ ^{6}$CPPM, Aix-Marseille Universit\'{e}, CNRS/IN2P3, Marseille, France\\
% ...
$ ^{9}$Fakult\"{a}t Physik, Technische Universit\"{a}t Dortmund, Dortmund, Germany\\
% ...
$ ^{37}$Universidad de Santiago de Compostela, Santiago de Compostela, Spain\\
$ ^{38}$European Organization for Nuclear Research (CERN), Geneva, Switzerland\\
% ...
$ ^{41}$Nikhef National Institute for Subatomic Physics, Amsterdam, The Netherlands\\
% ...
$ ^{46}$H.H. Wills Physics Laboratory, University of Bristol, Bristol, United Kingdom\\
% ...
$ ^{51}$School of Physics and Astronomy, University of Glasgow, Glasgow, United Kingdom\\
$ ^{52}$Oliver Lodge Laboratory, University of Liverpool, Liverpool, United Kingdom\\
% ...
}
\end{flushleft}

\end{document}